\documentclass[%
preprint,
 amsmath,amssymb,
 aps,
prb,
]{revtex4-2}

\usepackage{graphicx}% Include figure files
\usepackage{dcolumn}% Align table columns on decimal point
\usepackage{bm}% bold math
\usepackage{xcolor}
\usepackage{color,soul}

\begin{document}

%\preprint{APS/123-QED}

\title{Examining normal modes as fundamental heat carriers in amorphous solids: the case of amorphous silicon}% Force line breaks with \\

\author{Jaeyun Moon$^*$}
 \affiliation{Materials Science and Technology Division\\ Oak Ridge National Laboratory, Oak Ridge, TN 37831, USA}%Lines break automatically or can be forced with \\

\email{To whom correspondence should be addressed; E-mail:  moonj@ornl.gov}

\date{\today}% It is always \today, today,
Notice: This manuscript has been authored by UT-Battelle, LLC under Contract No. DE-AC05-00OR22725 with the U.S. Department of Energy. The United States Government retains and the publisher, by accepting the article for publication, acknowledges that the United States Government retains a non-exclusive, paid-up, irrevocable, world-wide license to publish or reproduce the published form of this manuscript, or allow others to do so, for United States Government purposes. The Department of Energy will provide public access to these results of federally sponsored research in accordance with the DOE Public Access Plan (http://energy.gov/downloads/doe-public-access-plan).

\begin{abstract}
Normal mode decomposition of atomic vibrations has been used to provide microscopic understanding of thermal transport in amorphous solids for decades. In normal mode methods, it is naturally assumed that atoms vibrate around their equilibrium positions and that individual normal modes are the fundamental vibrational excitations transporting heat. With the abundance of predictions from normal mode methods and experimental measurements now available, we carefully analyze these calculations in amorphous silicon, a model amorphous solid. We find a number of discrepancies, suggesting that treating individual normal modes as fundamental heat carriers may not be accurate in amorphous solids. Further, our classical and \textit{ab-initio} molecular dynamics simulations of amorphous silicon demonstrate a large degree of atomic diffusion, especially at high temperatures, leading to the conclusion that thermal transport in amorphous solids could be better described starting from the perspective of liquid dynamics rather than from crystalline solids.  

\end{abstract}

\maketitle

\clearpage
%\tableofcontents

\section{Introduction}
The atomic vibrations and thermal properties of amorphous dielectric solids are of fundamental and practical interest. For applications, amorphous solids are widely used as thermal insulators in thermopile and other detectors \cite{foote_thermopile_2003} where thermal shielding directly sets the sensitivity of the detector. Amorphous solids are of fundamental interest themselves because the lack of atomic periodicity complicates theoretical developments \cite{anderson_through_1995}. As a result, the nature of the vibrational excitations that carry heat in amorphous solids remains an active areas of research.

Historically, thermal transport in amorphous silicon (a-Si) has been studied extensively as a model amorphous solid due to its relatively simple monatomic composition and for its wide range of industrial applications such as solar cells \cite{carlson_amorphous_1976} and gravitational wave detectors \cite{birney_amorphous_2018}. Amorphous silicon gained further interests due to its anomalously strong thickness dependence in thermal conductivity measurements that are not observed in typical amorphous solids \cite{cahill_thermal_1994, liu_high_2009}. Allen and Feldman showed that their thermal conductivity model (now often referred to as diffuson thermal conductivity) could not capture the thermal conductivity of a-Si and ascribed this discrepancy to contribution from long mean free path, low frequency modes \cite{allen_thermal_1989}. Decomposing the atomic vibrations in amorphous silicon into individual normal modes, Allen and Feldman further proposed to categorize these into three types of quasi-particles: propagons, diffusons, and locons \cite{allen_diffusons_1999}. Propagons are propagating modes that can travel over distances larger than their corresponding wavelengths or interatomic distances. Locons are localized modes that are thought to contribute negligibly to thermal conductivity. Diffusons are modes that are neither propagating nor localized. Hence, the concepts of mean free paths and wavevectors lose meaning. For decades, numerous following works described the thermal transport in various amorphous solids including polymers \cite{lv_understanding_2017, kommandur_empirical_2017, feng_size_2020} and network glasses \cite{liao_akhiezer_2020, lv_phonon_2016} to amorphous nanocomposites \cite{moon_sub-amorphous_2016}, low dimensional amorphous solids \cite{zhu_phonons_2016}, and others \cite{zhou_contribution_2017} in terms of propagons, diffusons, and locons.  %While a few different methods have emerged to determine propagons versus diffusons over the years such as mode eigenvector periodicity and others \cite{seyf_method_2016, larkin_thermal_2014, lv_phonon_2016}, we strictly use the original criterion of Ioffe-Regel crossover in this work where the mean free path is smaller than the wavelength of the vibrational excitation proposed by Allen and Feldman \cite{allen_diffusons_1999}.

The literature on normal mode descriptions of a-Si has some variations in detailed conclusions, but the general consensus is that propagons contribute from 20 to 50 \% of overall thermal conductivity despite their small population consisting of only $\sim$ 3 \% of the density of states \cite{allen_diffusons_1999, he_heat_2011, larkin_thermal_2014, seyf_method_2016, lv_direct_2016, liao_akhiezer_2020, hashemi_effects_2020}, and that they are dominantly scattered by anharmonicity \cite{he_heat_2011, larkin_thermal_2014, liao_akhiezer_2020}. Here, anharmonicity refers to typical phonon-phonon interactions in the absence of disorder scattering \cite{he_heat_2011, larkin_thermal_2014} and this definition of anharmonicity is used in the current work unless otherwise specified. 

In this work, we carefully examine the conclusions from the rich body of literature on thermal transport in amorphous silicon that treat individual normal modes as fundamental heat carriers and find discrepancies on three levels: (i) the intrinsic assumption in normal modes that atoms vibrate around their equilibrium positions is not always valid, especially at high temperatures but still well below glass transition temperatures, (ii) conflicting predictions when comparing the normal mode results from one work to the other, and (iii) opposite temperature dependent trends in thermal conductivity when directly comparing with experiments. Our analysis, therefore, suggests that individual normal modes may not be accurate descriptors of fundamental heat carriers in amorphous solids. Rather than treating the fundamental heat carriers as normal modes in amorphous solids similar to crystals, it is our view that thermal transport in amorphous solids at all temperatures below the glass transition temperature should be more generally considered from the perspectives of liquid physics where one needs to also consider the effect of atomic diffusion in describing energy transport. 

This paper is organized as follows. In Section II, we describe normal modes and methods to characterize thermal transport including normal mode lifetimes, Green-Kubo modal analysis, and Allen-Feldman theory. In Section III, we point out discrepancies regarding the conclusions reached by these normal mode methods in studying thermal transport in amorphous solids. Finally, we provide some insights into how one might tackle this challenge more generally.

\section{Normal mode methods}

In normal mode analysis, the atomic displacements around an equilibrium position for atom $j$ in unit cell $l$ are expanded as plane wave solutions to the classical harmonic oscillator approximation as 
\begin{equation}
    \boldsymbol{u}(jl, t) = \sum_{\boldsymbol{k},\nu} \boldsymbol{U}(j, \boldsymbol{k}, \nu) e^{i[\boldsymbol{k} \cdot \boldsymbol{r}(jl) - \omega (\boldsymbol{k}, \nu)t]}
\end{equation}
where $\boldsymbol{k}$ is the wave vector, $\omega$ is the angular frequency, $\nu$ is the polarization, and $\boldsymbol{U}(j, \boldsymbol{k}, \nu)$ is the amplitude vector. Common to phonon transport literature, the above equation is re-written as \cite{dove_introduction_1993}
\begin{equation}
    \boldsymbol{u}(jl, t) = \frac{1}{(Nm_j)^{1/2}}\sum_{\boldsymbol{k},\nu} \boldsymbol{e}(j, \boldsymbol{k},\nu) e^{i\boldsymbol{k} \cdot \boldsymbol{r}(jl)} Q(\boldsymbol{k}, \nu, t)
    \label{distonor}
\end{equation}
where $N$ is the number of atoms, $\boldsymbol{e}(j, \boldsymbol{k},\nu)$ is the mode eigenvector, $m$ is the mass and $Q(\boldsymbol{k}, \nu, t)$ is the Fourier transform of (\ref{distonor}) as shown below 
\begin{equation}
    Q(\boldsymbol{k}, \nu, t) = \frac{1}{N^{1/2}} \sum_{jl} m_j^{1/2} e^{-i\boldsymbol{k} \cdot \boldsymbol{r}(jl)} \boldsymbol{e}^*(j, \boldsymbol{k},\nu) \cdot \boldsymbol{u}(jl, t)
\end{equation}
The complex quantity, $Q(\boldsymbol{k}, \nu, t)$, is named the normal mode coordinate. Many thermodynamic and dynamic properties such as heat capacity and root-mean-square displacements can then be decomposed in terms of these normal modes with the caveat that atoms vibrate around their equilibrium position and plane wave solutions are reasonable assumptions. More details about normal mode decomposition of such properties can be found in Refs. \cite{dove_introduction_1993} and \cite{mcgaughey_phonon_2006}. 

In subsequent subsections, three commonly used normal mode methods to characterize thermal transport such as normal mode lifetimes, Green-Kubo modal analysis, and Allen-Feldman theory are introduced in detail. In addition to the assumption that atoms vibrate around their equilibrium positions, these methods treat normal modes as fundamental heat carriers. 

\subsection{Normal mode lifetimes}

The Hamiltonian of a system of harmonic oscillators can be expressed equivalently in both real space and  normal mode space as follows
\begin{equation}
    H = \frac{1}{2} \sum_{jl} m_j | \boldsymbol{\dot u}(jl, t)|^2 + \frac{1}{2} \sum_{jj', ll'} \boldsymbol{u}^T (jl,t) \cdot \Phi(jj', ll') \cdot \boldsymbol{u}(j'l', t)
    \label{harmonic_H}
\end{equation}
and
\begin{equation}
    H = \frac{1}{2} \sum_{\boldsymbol{k}, \nu} \dot Q(\boldsymbol{k}, \nu, t) \dot Q(-\boldsymbol{k}, \nu, t) + \frac{1}{2} \sum_{\boldsymbol{k}, \nu} \omega^2 (\boldsymbol{k}, \nu) Q(\boldsymbol{k}, \nu, t) Q(-\boldsymbol{k}, \nu, t)
\end{equation}
where a dot over a variable represents a time derivative, superscript $T$ represent a transpose, and $\Phi(jj', ll')$ is the force constant matrix.  The first term and the second term in the above Hamiltonian expressions are kinetic energy and potential energy, respectively. The instantaneous, total energy of each mode of a classical system is then
\begin{equation}
    H_{\boldsymbol{k},\nu} =  \frac{1}{2} \dot Q(\boldsymbol{k}, \nu, t) \dot Q(-\boldsymbol{k}, \nu, t) + \frac{1}{2} \omega^2 (\boldsymbol{k}, \nu) Q(\boldsymbol{k}, \nu, t) Q(-\boldsymbol{k}, \nu, t)
\end{equation}
However, the harmonic expression above is not a good representation of systems at finite temperature because phonons can interact via anharmonicity in the interatomic potential. To account for the mode interactions, displacements from molecular dynamics (MD) at a desired finite temperature which account for all degrees of anharmonicities are used instead and projected to the normal mode coordinates, the temporal decay of the autocorrelation of $H_{\boldsymbol{k},\nu}$ is then related to the relaxation time of each mode by \cite{ladd_lattice_1986,mcgaughey_phonon_2006}
\begin{equation}
    \frac{H_{\boldsymbol{k},\nu}(t) H_{\boldsymbol{k},\nu} (0)}{H_{\boldsymbol{k},\nu}(0) H_{\boldsymbol{k},\nu}(0)} = e^{-\frac{t}{\tau_{\boldsymbol{k}, \nu}}}
\end{equation}

\begin{figure}
	\centering
	\includegraphics[width=0.8\linewidth]{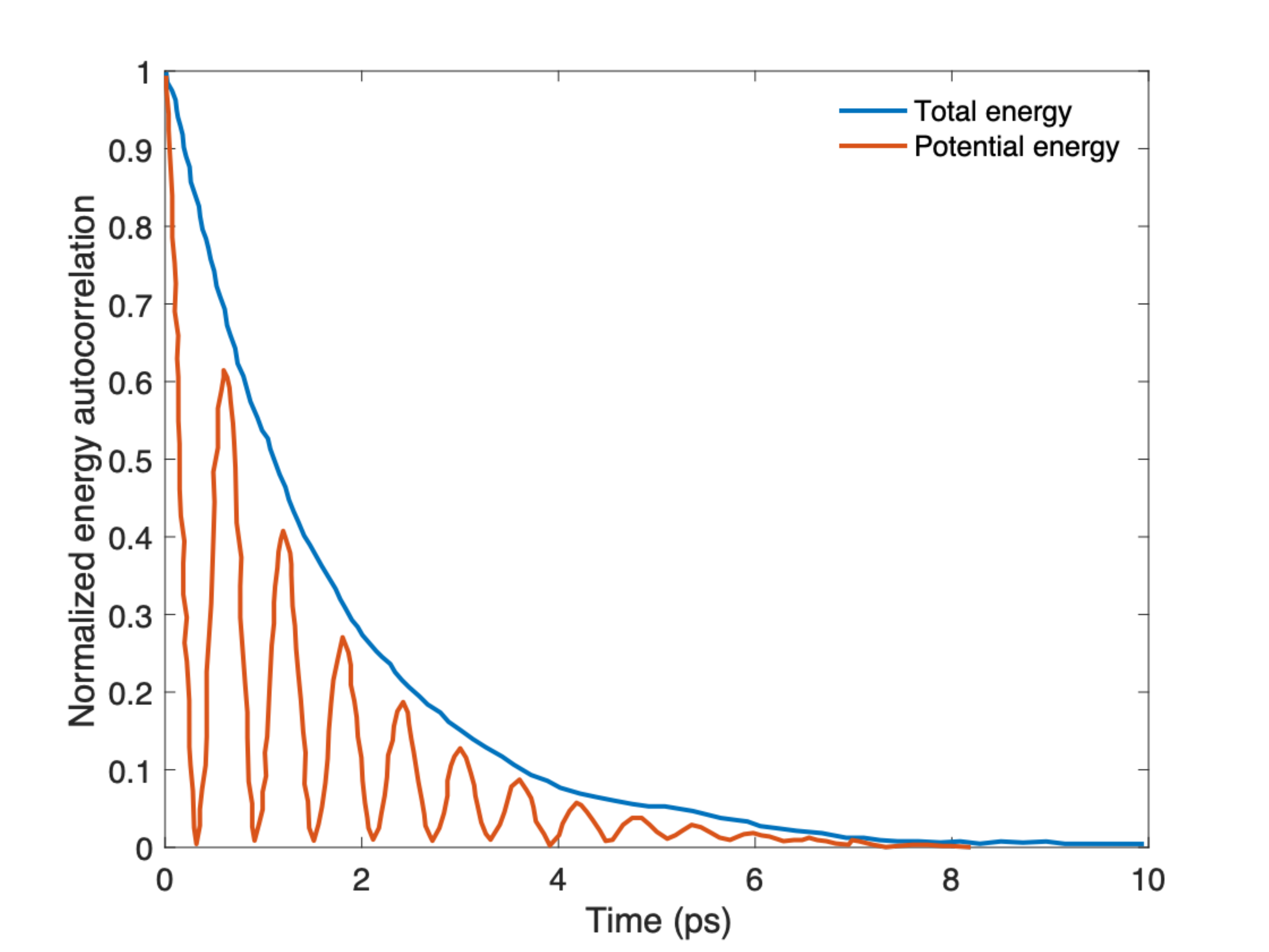}
	\caption{A typical normalized energy autocorrelation for the relaxation time calculations as adapted from \cite{mcgaughey_phonon_2006}. Blue and orange solid curves correspond to normalized total energy and potential energy autocorrelations of a mode, respectively. We see an exponential decay in the total energy autocorrelation as expected. The vibration frequency of the mode is one half of the oscillation frequency observed in the potential energy autocorrelations.  }
	\label{fig:lifetime_mcgaughey}
\end{figure}

A typical normalized total energy autocorrelation for a mode is shown in Fig. \ref{fig:lifetime_mcgaughey} \cite{mcgaughey_phonon_2006}. As expected, we see an exponential decay in the normalized energy autocorrelation and the lifetime can be extracted as discussed. Normalized potential energy autocorrelation is also plotted in the same figure. The vibration frequency of the mode is half of the oscillation frequency in the potential energy autocorrelation. The normal mode lifetime calculations can also be done in frequency space in which we see a Lorentzian function that determines the mode frequency (peak location) and mode lifetime (inverse of full width at half maximum). 

The normal mode lifetime calculation scheme mentioned above is general for both crystals and amorphous solids. However, it is important to note that for amorphous solids all the normal mode analysis is done at the $\Gamma (\boldsymbol{k} = 0)$ point due to lack of translational symmetry, which is equivalent to treating the entire computational domain as one supercell. Lifetimes obtained using the primitive cell as the unit cell and the simulation cell as the unit cell are nearly identical in crystals \cite{mcgaughey_predicting_2014}. When using the entire domain as the unit cell in crystals, symmetry operations can be performed to obtain correct modal group velocities. On the other hand, group velocities ($\frac{d \omega}{dk}$) of normal modes in amorphous solids are not directly found and some assumptions such as using the Debye sound velocity as the group velocities have to be made to calculate the thermal conductivity utilizing normal mode lifetimes \cite{larkin_thermal_2014, zhou_contribution_2017}. Propagon contribution to thermal conductivity is then calculated from the kinetic theory where the phonon particle picture is valid. The remaining diffuson thermal conductivity by Allen and Feldman theory as discussed in Section II C. is then added to obtain the overall thermal conductivity. 

Now that the essence of the normal mode lifetime calculations is introduced, we next explore the Green-Kubo modal analysis (GKMA). 

\subsection{Green-Kubo modal analysis}
Green-Kubo modal analysis is a method to decompose the heat flux into the time derivative of normal mode coordinates in the Green-Kubo formalism to calculate the thermal conductivity \cite{lv_direct_2016}. According to the Green-Kubo formalism, the thermal conductivity tensor is given by
\begin{equation}
    k_{\alpha\beta} = \frac{V}{k_B T^2} \int \langle J_{\alpha} (t + t') J_{\beta} (t) \rangle dt'
    \label{Green-Kubo}
\end{equation}
where the angled bracket notation is the ensemble average, subscripts $\alpha$ and $\beta$ denote Cartesian directions, $V$ is the volume of the system, $k_B$ is the Boltzmann constant, $T$ is the temperature, and $J$ is the heat flux expressed as \cite{hardy_energy-flux_1963}
\begin{equation}
    \boldsymbol{J} = \frac{1}{V} \bigg[ \sum_i E_i \boldsymbol{v}_i + \frac{1}{2} \sum_{i,j} (\boldsymbol{F}_{ij} \cdot \boldsymbol{v}_i ) \boldsymbol{r}_{ij} \bigg]
    \label{hardy}
\end{equation}
where the summation is over atoms in the defined volume. It is worth mentioning that sometimes in the literature, the Green-Kubo thermal conductivity is written in terms of heat current, $\boldsymbol{S} = V\boldsymbol{J}$. So far, the Green-Kubo thermal conductivity with the above heat flux equation is general in that it can be used to calculate thermal conductivity of solids, liquids, and gases. Assuming atoms vibrating with respect to their equilibrium positions, velocity in (\ref{hardy}) is decomposed into time derivatives of normal modes as \cite{lv_direct_2016}
\begin{equation}
    \boldsymbol{v}_i (t) = \sum_n \boldsymbol{v}_i (n, t) = \frac{1}{m_i^{1/2}} \sum_n \boldsymbol{e} (i, n) \dot Q(n,t)
    \label{velocity_n}
\end{equation}
where the sum is over all modes. To reach this expression, we have treated the entire computational domain as one supercell as mentioned before.
Substituting (\ref{velocity_n}) into (\ref{hardy}), we obtain the individual modal contribution to the heat flux as
\begin{equation}
    \boldsymbol{J}(n, t) = \frac{1}{V} \bigg[ \sum_i E_i \bigg ( \frac{1}{m_i^{1/2}} \boldsymbol{e} (i, n) \dot Q(n,t) \bigg ) + \frac{1}{2} \sum_{i,j} \bigg \{ \boldsymbol{F}_{ij} \cdot \bigg ( \frac{1}{m_i^{1/2}} \boldsymbol{e} (i, n) \dot Q(n,t) \bigg ) \bigg \} \boldsymbol{r}_{ij} \bigg]
\end{equation}
Decomposition of the heat flux into individual mode contributions enables mode resolved thermal conductivity as
\begin{equation}
    k_{\alpha \beta, nn'} = \frac{V}{k_BT^2} \int \langle J_{\alpha}(n, t+t') J_{\beta}(n', t) \rangle dt'
\end{equation}
One can, therefore, examine how the correlation between pairs of modes contributes to thermal conductivity. The total thermal conductivity is then
\begin{equation}
    k_{\alpha \beta} = \frac{V}{k_BT^2} \sum_{n, n'} \int \langle J_{\alpha}(n, t+t') J_{\beta}(n', t) \rangle dt'
\end{equation}

Thermal conductivity calculations based on GKMA account for all degrees of anharmonicity as the velocities which are decomposed into normal modes are from classical molecular dynamics. Since the spectral thermal conductivity is known in this method, quantum correction can be made to the spectral specific heat and thus thermal conductivity. 

\subsection{Allen and Feldman theory}
Allen and Feldman proposed a quantum mechanical theory in which heat is carried by decoupled harmonic oscillators \cite{allen_thermal_1989, allen_thermal_1993}. Extensive derivation of the theory is well-documented in Ref. \cite{allen_thermal_1993}. Hence, only the important aspects of the derivations are covered here. 

The heat flux is written with respect to the heat flux operator as 
\begin{equation}
    \boldsymbol{J} = tr(\rho \boldsymbol{S}) = - \boldsymbol{k}\cdot \nabla T
\end{equation}
where $tr$ is the trace. Assuming a system in steady state with a local space-dependent temperature $T(x) = [k_B \beta (s)]^{-1}$, the local density matrix is given by
\begin{equation}
    \rho = \frac{e^{-\int d^3x \beta(x) h(x)}}{Z}
\end{equation}
where $h(x)$ is the Hamiltonian density operator and $Z$ is the partition function. The Hamiltonian can, therefore, be written in terms of $h(x)$ as $H = \int d^3 x h(x)$. Under the harmonic approximation, the Hamiltonian is equivalent to (\ref{harmonic_H}). The Hamiltonian density operator, $h(x)$ and the heat flux density operator $\boldsymbol{S}(x)$  obey the condition of local energy conservation as 
\begin{equation}
    \frac{\partial h(x)}{\partial t} + \nabla \cdot \boldsymbol{S}(x) = 0
\end{equation}
Now, recalling that the heat flux $\boldsymbol{J}$ vanishes in equilibrium when $\beta(x)$ is constant and assuming that the temperature fluctuations, $\delta T(x)$, are small, $\beta (x)$ can be expanded as
\begin{equation}
    \beta(x) = \beta \Big ( 1-\frac{\delta T}{T} \Big)
\end{equation}
where $\beta$ and $T$ are corresponding average constants. Substituting the expression back to the local density matrix, 
\begin{equation}
    \rho = \frac{e^{- \beta (H + H')}}{Z}
\end{equation}
\begin{equation}
    H' = -\frac{1}{T} \int d^3 x \delta T(x) h(x)
\end{equation}
Using the energy conversion condition to replace $h(x)$ by $\boldsymbol{S} = \frac{1}{V} \int d^3 x \boldsymbol{S}(x)$ and with some algebra, 
\begin{equation}
    H' = - \frac{1}{T} \int_{-\infty}^0 dt \int d^3x \nabla T(x) \cdot \boldsymbol{S}(t)
\end{equation}
 
Taking $\nabla T(x)$ as a constant, 
\begin{equation}
    H' = -\frac{V \nabla T}{T} \int_{-\infty}^{0}dt  \boldsymbol{S}
\end{equation}
Recognizing that the density matrix can be expanded in powers of the Hamiltonian perturbation,
\begin{equation}
    \rho = \frac{e^{- \beta (H + H')}}{Z} = \frac{e^{-\beta H}}{Z} \bigg ( 1 + \int_0^{\beta} d\lambda e^{\lambda H} H' e^{-\lambda H} + ... \bigg )
\end{equation}
When carrying out $ tr(\rho \boldsymbol{S})$, the first term is zero due to constant $\beta$ and the perturbation from the equation above leads to $-\boldsymbol{k}\cdot \nabla T$. The thermal conductivity can, therefore, be written as
\begin{equation}
    k_{\alpha\beta} = \frac{V}{T}\int_0^{\beta}d\lambda \int_0^{\infty} \langle e^{\lambda H} S_{\alpha}(t) e^{-\lambda H} S_{\beta}(0) \rangle
\end{equation}
Time shifting the time integral and taking the Fourier transform of the above equation, we obtain
\begin{equation}
    k_{\alpha\beta} = \frac{V}{T}\int_0^{\beta}d\lambda \int_0^{\infty} e^{i(\omega + i\eta)t}\langle S_{\alpha}(-i\hbar \lambda)  S_{\beta}(t) \rangle
\end{equation}
where $S(-i\hbar \lambda) \equiv e^{\lambda H}S e^{-\lambda H}$. No assumption of normal modes has been made so far and the above equation is the Kubo formula for thermal conductivity. Decomposing the heat current operator in terms of normal modes in the harmonic approximation, expressing the time dependence of the heat current operator in the Heisenberg picture, and taking the real part of the thermal conductivity, we obtain
\begin{equation}
    k = \frac{1}{V}\sum_i C_i(T) D_i
    \label{k_AF}
\end{equation}
where $C_i(T)$ is the specific heat and $D_i$ is the mode diffusivity given by
\begin{equation}
    D_i = \frac{\pi V^2}{3\hbar \omega_i^2} \sum_{i \neq j} |S_{ij}|^2 \delta(\omega_i - \omega_j)
\end{equation}
The matrix elements of the heat current operator are given by
\begin{equation}
\begin{split}
    S_{ij} &= \frac{\hbar}{2V}\boldsymbol{v}_{\boldsymbol{K}ij}(\omega_{\boldsymbol{K}i} + \omega_{\boldsymbol{K}j})\\
    \boldsymbol{v}_{\boldsymbol{K}ij} &= \frac{i}{2\sqrt{\omega_{\boldsymbol{K}i}\omega_{\boldsymbol{K}j}}}\sum_{\alpha, \beta, m, \kappa, \kappa'} e_{\alpha}(\kappa;\boldsymbol{K},i)D_{\beta\alpha}^{\kappa' \kappa}(0,m) (\boldsymbol{R}_m + \boldsymbol{R}_{\kappa \kappa'})e^{i\boldsymbol{K}\cdot\boldsymbol{R}_m}  e_{\beta}(\kappa';\boldsymbol{K},j)
\end{split}
\end{equation}
where $\boldsymbol{K}$ is the wavevector, $\alpha, \beta$ are the Cartesian directions, $m$ labels a unit cell, $\kappa, \kappa'$ denote the atoms in the cell $m$, $e$ with subscripts is the eigenvector, $D_{\beta\alpha}^{\kappa' \kappa}(0,m)$ is the Hermitian force constant matrix. 
The above expression for $ \boldsymbol{v}_{\boldsymbol{K}ij}$ is written for a periodic system but can be simplified to amorphous solids by considering the entire domain as one unit cell. 

The above thermal conductivity expression in Eq. \ref{k_AF} is termed the Allen-Feldman thermal conductivity also denoted as $k_{AF}$. To summarize the derivations, the temperature gradient in a disordered solid couples different harmonic eigenstates through the heat current operator (off-diagonal elements) which lead to finite thermal conductivity while the diagonal elements contribute zero to the overall thermal conductivity confirming that the thermal conductivity using the above expressions is from non-propagating modes. 

With these three widely used normal mode methods reviewed, we reiterate here what the normal mode methods predict of thermal transport in amorphous silicon \cite{larkin_thermal_2014, he_heat_2011, seyf_method_2016, lv_direct_2016, hashemi_effects_2020, liao_akhiezer_2020}. Propagons exist up to 2 to 3 THz, consisting of only a few \% of the density of states. Due to their long mean free paths, however, they can still contribute a significant amount (20 \% to 50 \%) to the total thermal conduction and are expected to be insensitive to disorder, suggested from $\tau \sim \omega^{-2}$ trends while the rest of the thermal conductivity is from diffusons. 

\section{Discrepancies found in the predictions from normal mode methods}

\subsection{Prediction: propagons exist up to 2 to 3 THz in a-Si.} 
\begin{figure}
	\centering
	\includegraphics[width=0.85\linewidth]{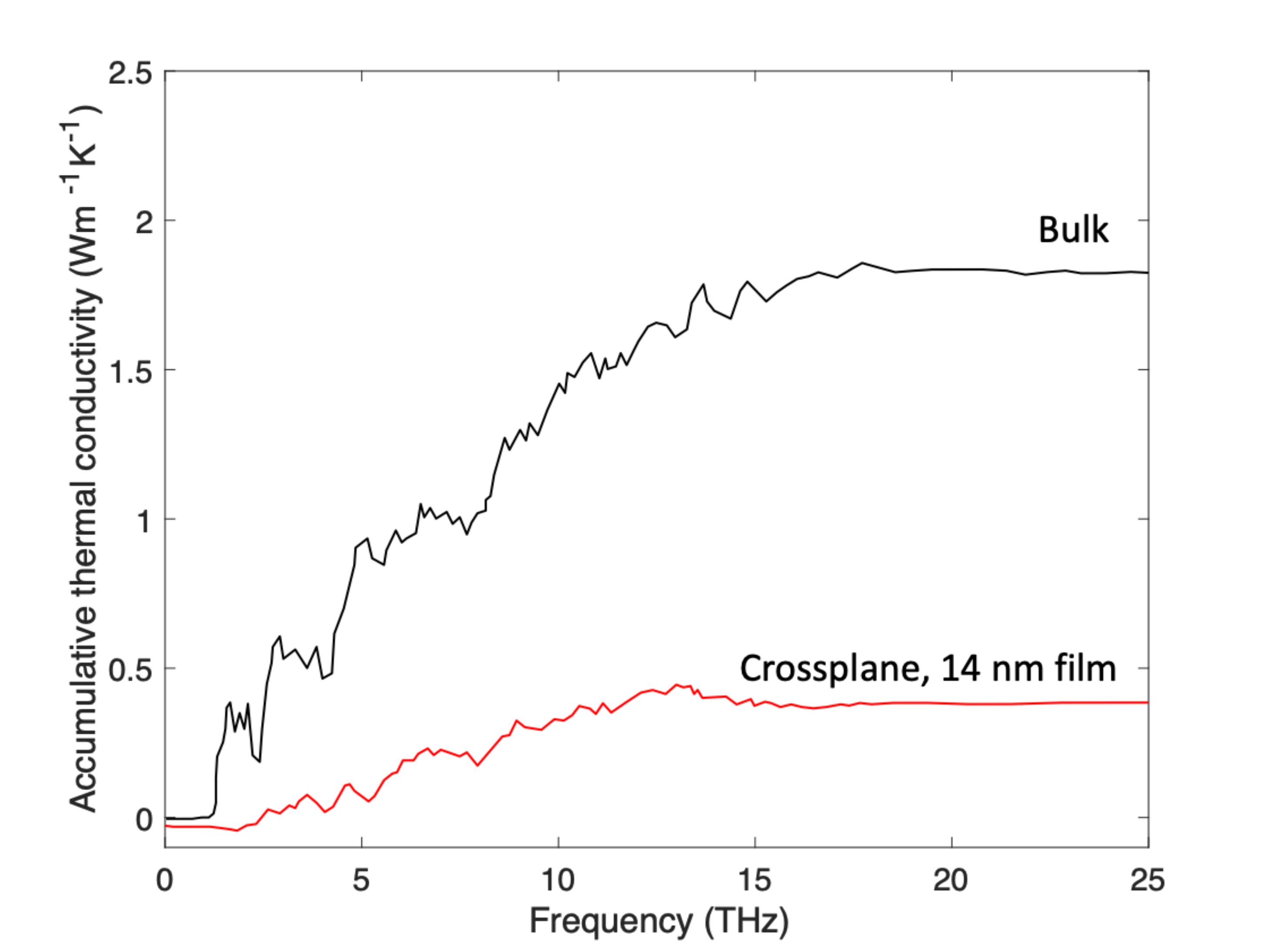}
	\caption{Accumulative spectral thermal conductivity of bulk a-Si (solid black line) and 14 nm a-Si thin film (solid red line) at 300 K using GKMA  \cite{deangelis_thermal_2018}. A clear suppression in the thermal conductivity from both propagons and diffusons is seen.}
	\label{fig:GKMA_size}
\end{figure}

We first examine the general consensus that propagons exist up to 2 to 3 THz. If this assertion is true, we should not observe large size effects in the thermal transport by diffusons that are present from $\sim$ 2 THz to $\sim$ 17 THz by the definition of propagons and diffusons. In our prior work, GKMA was utilized to study the size effect of a-Si by comparing the spectral thermal conductivity of bulk a-Si \cite{lv_direct_2016} and a 14 nm thick a-Si film \cite{deangelis_thermal_2018} as shown in Fig. \ref{fig:GKMA_size}. To clarify, bulk a-Si here denotes the 4096-atom domain with the cube length of around 4.5 nm with periodic boundary conditions in all Cartesian directions as in Ref. \cite{lv_direct_2016}. Tersoff potential was used. Size effects for this "bulk" structure are still expected for long-wavelength and low frequency vibrations as demonstrated below 1 THz. The 14 nm film has dimensions of 14 nm by 2 nm by 2 nm. In the long direction (cross-plane), the structure is exposed to vacuum while in the in-plane directions, periodic boundary conditions are imposed. It is worth mentioning that 14 nm is about two orders of magnitude larger than the interatomic distance of a-Si which is $\sim$ 2.3 \AA. A stark difference in the spectral thermal conductivity of both structures is observed. A noticeable suppression of the propagon thermal conductivity below 2 $\sim$ 3 THz is observed in the cross-plane direction as expected. What is unexpected here is the clear suppression of diffuson thermal conductivity. By the definition of diffusons by Allen and Feldman \cite{allen_diffusons_1999}, they are normal modes with ill-defined mean free paths below distances on the order of interatomic distances (a few \AA). Hence, if the normal modes from 2 THz to 17 THz are indeed diffusons, the diffuson spectral thermal conductivity should not be affected in the cross-plane direction of the 14 nm thick a-Si structure. %A clear discrepancy is, therefore, observed here. 

\subsection{Prediction: propagons are dominantly scattered by anharmonicity rather than disorder}
We next discuss the prediction that the lifetimes of few THz vibrations are governed by anharmonicity \cite{he_heat_2011, larkin_thermal_2014, liao_akhiezer_2020}.  If that is the case, explaining the low thermal conductivity of a-Si is challenging because the same vibrations contribute 75 Wm\textsuperscript{-1}K\textsuperscript{-1} to thermal conductivity in c-Si. Accounting for the low thermal conductivity of a-Si only by changes in anharmonicity requires either unphysically large increases in anharmonic force constants or in the scattering phase space. These changes would in turn affect other properties like the heat capacity of a-Si that have not been observed \cite{zink_thermal_2006}. Along similar lines, if lifetimes of few THz vibrations are governed by anharmonicity the reported thermal conductivities of films of the same thickness should be reasonably uniform, yet the data vary widely \cite{cahill_thermal_1994,zink_thermal_2006,liu_high_2009}.

\begin{figure}
	\centering
	\includegraphics[width=0.77\linewidth]{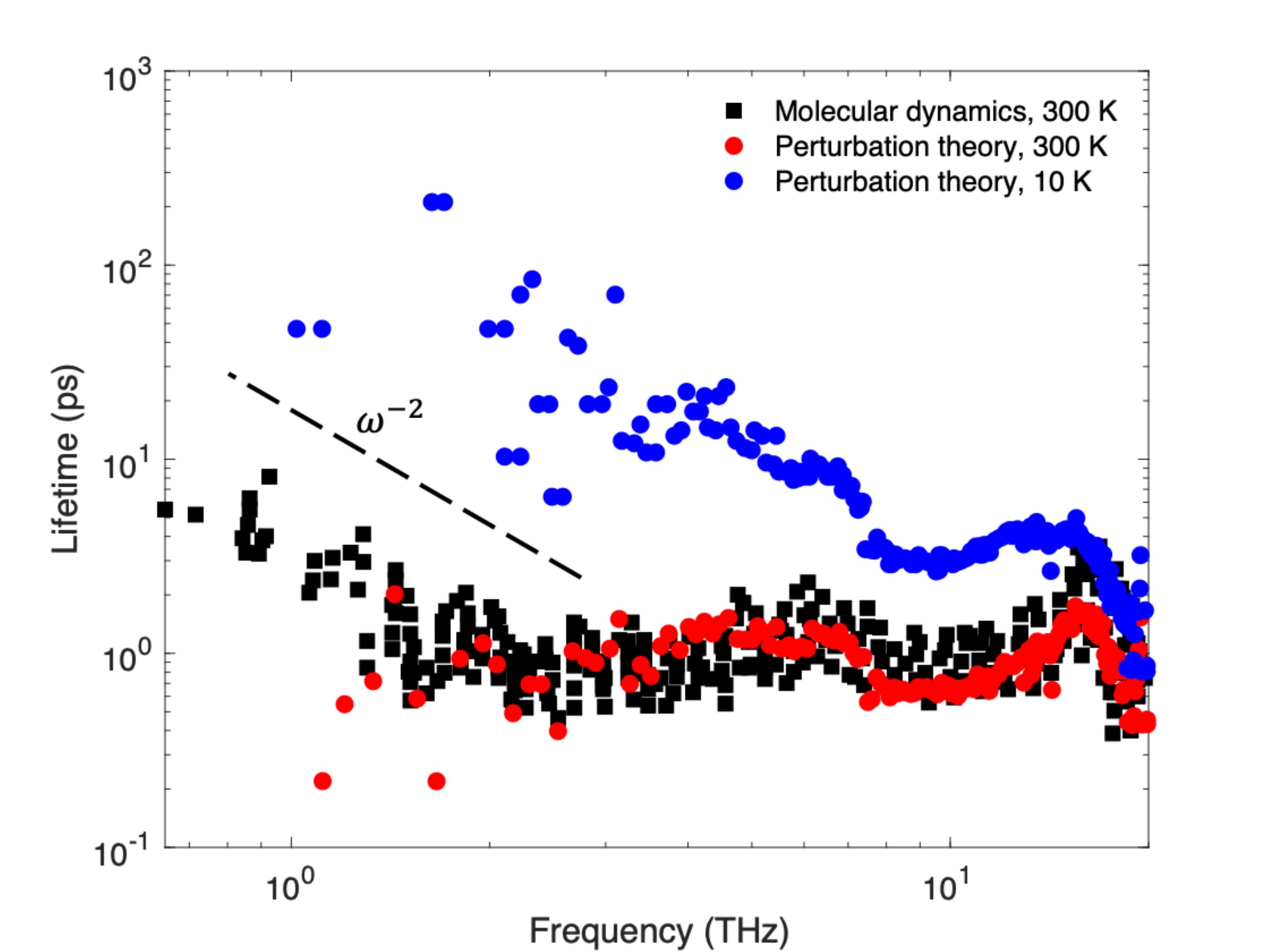}
	\caption{Normal mode lifetimes calculated from molecular dynamics at 300 K (black solid squares) \cite{larkin_thermal_2014} and perturbation theory at 10 K (blue solid circles) and 300 K (red solid circle) \cite{fabian_numerical_2003}. For all cases, the Stillinger-Weber potential \cite{stillinger_computer_1985} was used and WWW continuous random network algorithms were used to construct the structure. At 300 K, molecular dynamics and perturbation theory calculations produce consistent lifetimes and at 10 K, one to two orders of magnitude increase in the lifetimes are observed for modes with frequency less than 5 THz. $\omega ^{-2}$ is plotted as a guide to the eye, which signifies the anharmonic scattering \cite{larkin_thermal_2014, callaway_model_1959}.  }
	\label{fig:normal_mode_lifetime}
\end{figure}

To elucidate the discrepancy further, some of the normal mode lifetimes evaluated at $\Gamma(\boldsymbol{k}=0)$ in amorphous silicon (a-Si) from several research groups spanning a couple decades \cite{fabian_anharmonic_1996, bickham_calculation_1998, fabian_numerical_2003, he_heat_2011, larkin_thermal_2014, lv_examining_2016} are illustrated in Fig. \ref{fig:normal_mode_lifetime}. In this figure, lifetimes from molecular dynamics \cite{larkin_thermal_2014} and from perturbation theory \cite{fabian_anharmonic_1996,fabian_numerical_2003} that use the Stillinger-Weber potential \cite{stillinger_computer_1985} are representatively shown.  %Some works further calculated modal thermal conductivity of a-Si using these normal mode lifetimes by calculating modal group velocities \cite{he_heat_2011} or by making the assumption that their group velocity is the Debye sound velocity \cite{larkin_thermal_2014}. However, it is not clear how modal group velocities were calculated as the lifetimes were strictly calculated at $\boldsymbol{\Gamma} (\boldsymbol{k} = 0)$ by treating the entire structure as a supercell. Each of these works has used a different interatomic potential (Stillinger-Weber \cite{stillinger_computer_1985} vs. Tersoff \cite{tersoff_modeling_1989}) and made the structures differently (continuous random network vs. melt-quench) but their results are consistent \cite{he_heat_2011, larkin_thermal_2014}. They have concluded that while propagating waves exist only up to 2 THz, they contribute a substantial portion of the total thermal conductivity (40 to 50 \%) due to their large mean free paths. Another finding was that these propagating waves are not sensitive to the disorder and are instead scattered by anharmonic interactions with one another \cite{he_heat_2011, larkin_thermal_2014}. However, there are some discrepancies with these predictions from the normal mode lifetime calculations as discussed more in detail below.
$\omega ^{-2}$ scaling is plotted as a guide to the eye, which signifies the anharmonic scattering \cite{larkin_thermal_2014, callaway_model_1959}.  %and the structures (4096 atoms for MD and 1000 atoms for PT) were generated from Wooten-Winer-Weaire (WWW) continuous random network algorithms. 
We see that at 300 K, the lifetimes derived from both methods coincide well with the exception of a couple points at low frequencies using the perturbation theory method. At 10 K, one to two orders of magnitude increase in the lifetimes for frequencies below around 5 THz is clearly observed, which is consistent with their conclusion that scattering of propagating waves are governed by anharmonicity, unaffected by the disorder. This prediction, therefore, indicates that thermal diffusivity and conductivity should necessarily have a  significant and continuous increase from 300 K to 10 K. Taking into account the temperature dependent phonon occupation and assuming that the lifetimes increase on average by a factor of 100 at these frequencies and that normal mode group velocities are sound velocities, one can estimate that the propagon thermal conductivity increases by a factor of 40 from 300 K to 10 K, resulting in the overall thermal conductivity from 1.7 to 23 Wm\textsuperscript{-1}K\textsuperscript{-1}. However, a large increase in thermal conductivity going from room temperature to low temperatures on the order of 10 K is not observed in several experiments \cite{cahill_thermal_1994, liu_high_2009, kwon_unusually_2017}. Rather, the opposite trend closely following specific heat temperature dependence is apparent. Our recent picosecond acoustics mean free path measurements of 100 GHz excitations in amorphous silicon further show that the macroscopic mean free paths exceeding 10 $\mu m$ increase only slightly by 60\% to 70\% from room temperature to $\sim$ 40 K rather than a factor of $\sim$ 100 \cite{kim_origin_2021}.

One may think that the discrepancy in the normal mode lifetime works mentioned above may be the result of using unphysically anharmonic interatomic potentials rather than the normal mode lifetime methodology itself. However, using the same interatomic potentials, we find that the thermal conductivity from the Green-Kubo formalism follows a typical glass temperature dependent thermal conductivity, which signifies that the strong anharmonicity found in the normal mode lifetimes does not originate from the interatomic potentials themselves \cite{moon_sub-amorphous_2016}. 

Our hypothesis is that the discrepancy stems from obtaining the normal modes by treating the entire domain as the unit cell. In doing so, the structural disorder information (a weak function of temperature) is lost as the dynamical matrix is evaluated for the entire domain and the only way the normal modes are scattered is, then, through interacting with other normal modes, leading to the often observed $\tau \sim \omega^{-2}$ which is also demonstrated in other amorphous solids \cite{zhou_contribution_2017, lv_phonon_2016}. We will investigate this more in detail in our future works.  

It is worth noting here that at low temperatures typically below 1 K, phenomenological two level system (TLS) model is successful in describing temperature dependent thermal conductivity and heat capacity of amorphous solids \mbox{\cite{phillips_tunneling_1972, anderson_anomalous_1972}} in which two level system states further scatter normal modes. Several following works have included TLS scattering with various fitting parameters in calculating spectral diffusivity and could, in turn, reproduce the experimental temperature dependent thermal conductivity of amorphous silicon   \cite{sheng_heat_1991, feldman_thermal_1993,graebner_phonon_1986, kwon_unusually_2017}. However, amorphous silicon with near total absence of TLS characterized by internal friction measurements still exhibited the typical glass-like thermal conductivity temperature dependence, suggesting that including the TLS model to induce the glass-like temperature dependence of thermal conductivity may not be justified  \cite{liu_high_2009, zink_thermal_2006}. There have also been some questions to the uniqueness of the phenomenological TLS model to describe the apparently successful low temperature dependent heat capacity and thermal conductivity \cite{leggett_tunneling_2013}.

\subsection{Assumption: atoms move around equilibrium positions in glasses}
Here, we examine the fundamental assumption of normal mode analysis that atoms vibrate around their equilibrium positions.

Two types of amorphous silicon simulation domains were prepared in this work. The first structure (4096 atoms) was created using the Stillinger Weber potential \cite{stillinger_computer_1985} and a timestep of 0.5 fs in classical molecular dynamics (LAMMPS \cite{plimpton_fast_1995}). The crystalline structure was first melted at 3500 K for 500 ps in an NVT ensemble. Next, the liquid silicon was quenched to 1000 K with the quench rate of 100 K  ps\textsuperscript{-1}. The structures were annealed at 1000 K for 25 ns to reduce metastabilities. Finally, the domain was quenched at a rate of 100 K ps\textsuperscript{-1} to desired temperatures (300 K, 500 K, and 900 K) and equilibrated at these temperatures for 10 ns in an NPT ensemble using a Nose-Hoover thermostat. Similar melt-quench-anneal processes have also been utilized in several prior amorphous silicon works \cite{larkin_thermal_2014, lv_direct_2016, moon_propagating_2018}. The atomic locations were then collected for the next 150 ps.

\begin{figure}
	\centering
	\includegraphics[width=0.85\linewidth]{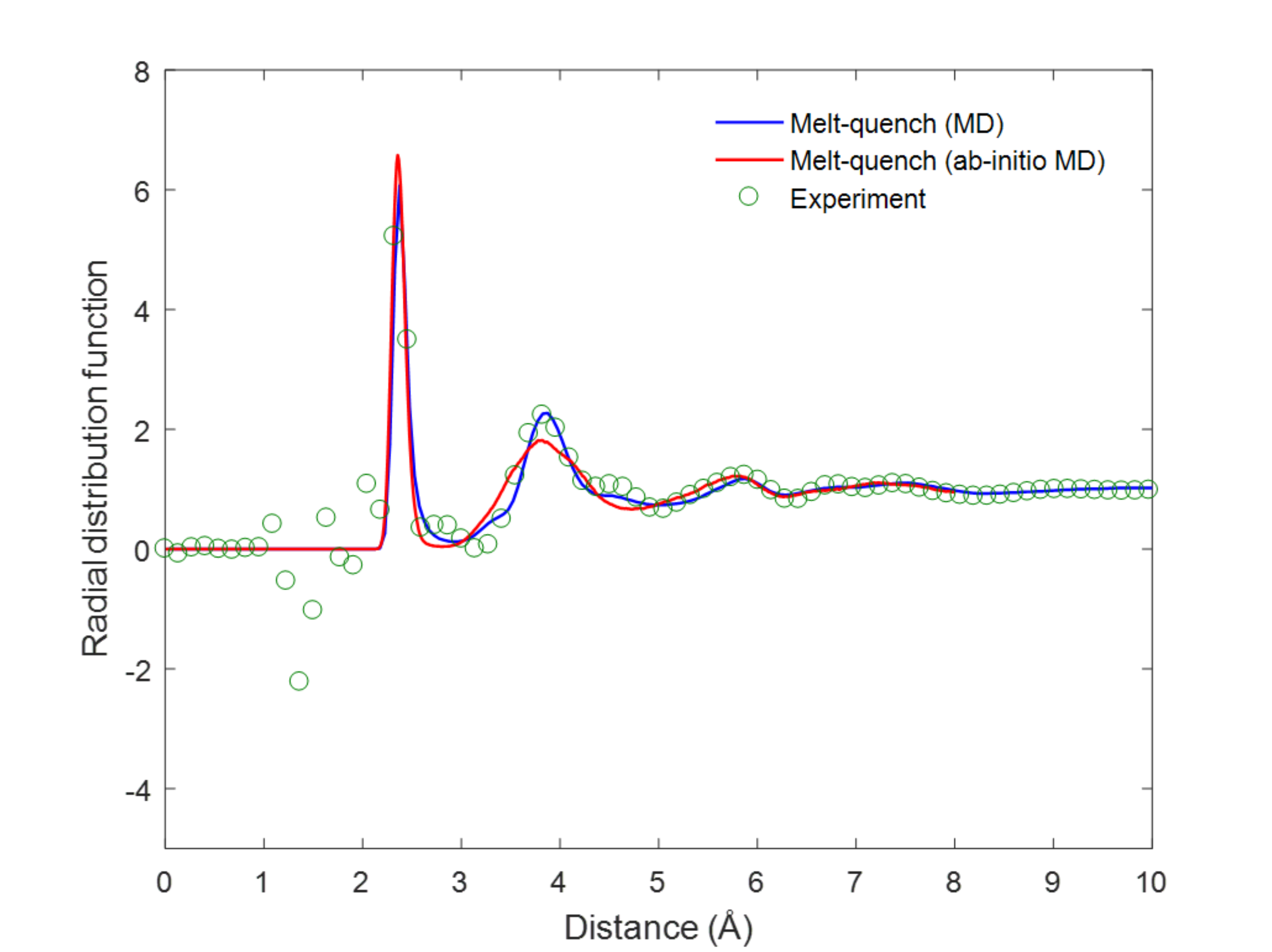}
	\caption{Radial distribution function of two a-Si structures made from MD and \textit{ab-initio} MD is compared to neutron diffraction of a-Si from our previous work \cite{moon_thermal_2019}.}
	\label{fig:RDF_ab_initio}
\end{figure}

Another structure (216 atoms) was also created by \textit{ab-initio} molecular dynamics (AIMD). AIMD simulations were performed with the projector augmented wave approach \cite{blochl_first-principles_1993} and the Perdew-Burke-Ernzerhof (PBE) \cite{perdew_generalized_1996} generalized-gradient approximation to the density functional theory (DFT) as implemented in the Vienna ab-initio simulation package (VASP) \cite{kresse_efficient_1996}. The time step was the same as the classical MD value. Only $\Gamma$ was adopted for the wavevector grid  and the energy cutoff of 245 eV was used. Crystalline silicon (c-Si) was first melted at 3500 K for 30 ps followed by quenching to desired temperatures at 100 K/ps. The structure relaxed for 10 ps. Then, the atomic locations were recorded for the next 25 ps at the desired temperatures. The radial distribution function of these two structures compared to neutron diffraction measurements on a-Si \cite{moon_thermal_2019} is shown in Fig. \ref{fig:RDF_ab_initio}. Generally, good agreement between the computational structures and the experiment is observed, but a better match is demonstrated for the classical MD structure, especially at the second peak. Bond angle distribution functions for both structures demonstrate the average angle to be $\sim$ 109$^\circ$ with full width at half maximum of about 20$^\circ$ at room temperature, consistent with prior amorphous silicon MD works \cite{luedtke_preparation_1989, barkema_high-quality_2000}. 

\begin{figure} 
	\centering
	\includegraphics[width=0.52\linewidth]{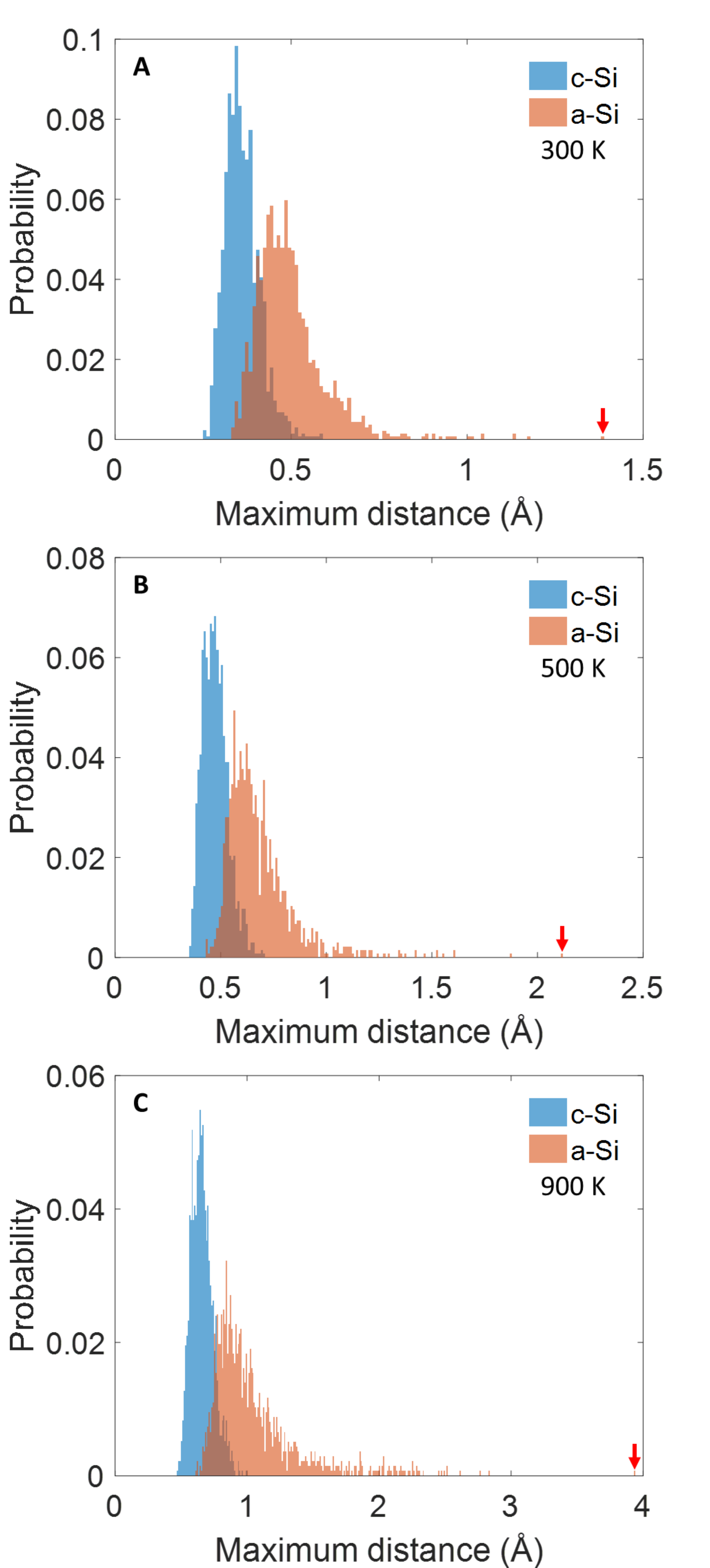}
	\caption{Histogram of maximum distance from initial position of each atom for both c-Si and a-Si from classical MD at (\textbf{A}) 300 K, (\textbf{B}) 500 K, and (\textbf{C}) 900 K. Red arrows denote the maximum value to which the distribution is extended to. We see that atoms in a-Si have consistently larger displacements than those in c-Si.}
	\label{fig:max_disp}
\end{figure}

\begin{figure}
	\centering
	\includegraphics[width=0.8\linewidth]{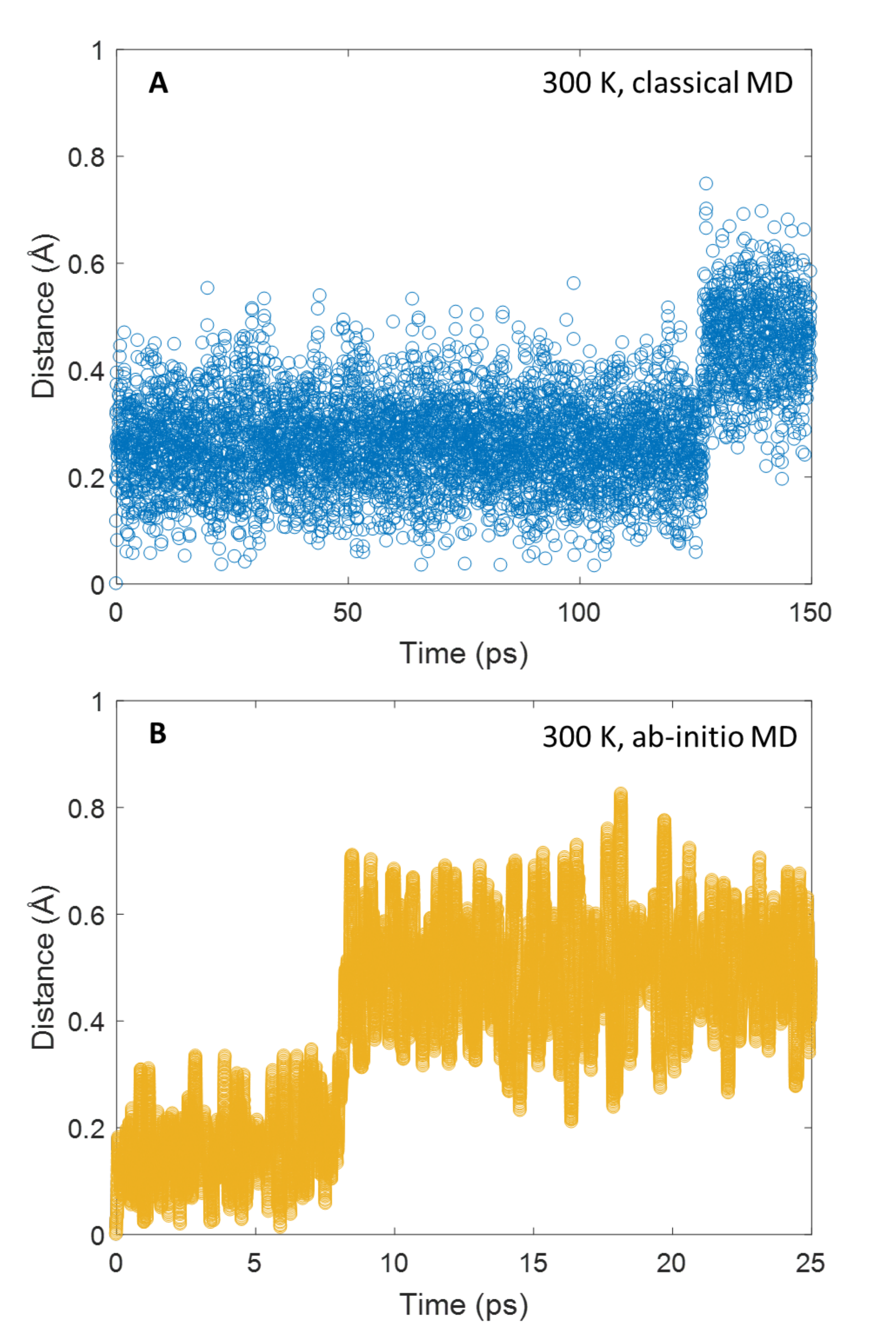}
	\caption{Temporal movement of an atom at 300 K for (\textbf{A}) classical MD and (\textbf{B}) ab-initio MD. These atoms have maximum distance of 0.7 to 0.8 \AA \, shown in Fig. \ref{fig:max_disp} (\textbf{A}). Clear atomic hopping is observed in both cases.}
	\label{fig:atom_jump}
\end{figure}

The maximum distances away from the initial positions for each atom during the 150 ps data collection for both c-Si and a-Si in classical MD are plotted in Fig. \ref{fig:max_disp}. It is clear that at all temperatures, atoms in a-Si explore significantly more volume compared to those in c-Si during the simulation. While having more volume to move around does not necessarily mean that atoms are diffusing, some atoms move as much as 1.4 \AA, 2.1 \AA, and 3.9 \AA \, which is comparable to the interatomic distance ($\sim$ 2.4 \AA) of a-Si at 300 K, 500 K, and 900 K, respectively. For simplicity, the maximum atomic distances from \textit{ab-initio} MD are not plotted as they had closely overlapping results with the classical MD calculations. The maximum distances for the \textit{ab-initio} MD calculations for 300 K, 500 K, and 900 K were 1.6 \AA, 2.5 \AA, and 4.3 \AA, respectively. To gain further insight into atomic displacements, temporal movement of an atom at 300 K having maximum distance of 0.7 to 0.8 \AA \ (smaller than the interatomic distance of $\sim$ 2.4 \AA) away from the initial position in Fig. \ref{fig:max_disp} \textbf{A} is shown in Fig. \ref{fig:atom_jump}. Also plotted is the results from \textit{ab-initio} MD. We can clearly see that in both cases, atoms initially vibrate around their equilibrium positions. Some time later, however, these atoms find new equilibrium positions and vibrate around these positions. When examining the temporal movements of atoms in a-Si at higher temperatures, more atoms were found to transition to new equilibrium positions due to their higher kinetic energy, as expected. For c-Si, atoms had greater distance amplitude at higher temperatures but they still vibrated around their equilibrium positions during the entire simulations. For amorphous silicon, the atomic displacement into the new equilibrium position shown in Fig. \ref{fig:atom_jump} is around 0.3 \AA, well below the interatomic distance of $\sim$ 2.4 \AA. In liquids and glasses, atomic diffusion processes can differ from crystals: due to the lack of fixed lattice, neighboring atoms can also readily rearrange, allowing atomic diffusion with short displacements in contrast to typical atomic hopping into vacancies and other defects in crystals. 

Atomic diffusion in the time scale of 1 to 100 ps means that propagons that are reported to have minimum normal mode lifetimes of 10 ps \cite{he_heat_2011} at room temperature will be further scattered by atomic diffusion which is not included in normal mode lifetime calculations as the atomic vibrations are projected to eigenvectors at 0 K.

\begin{figure}
	\centering
	\includegraphics[width=0.75\linewidth]{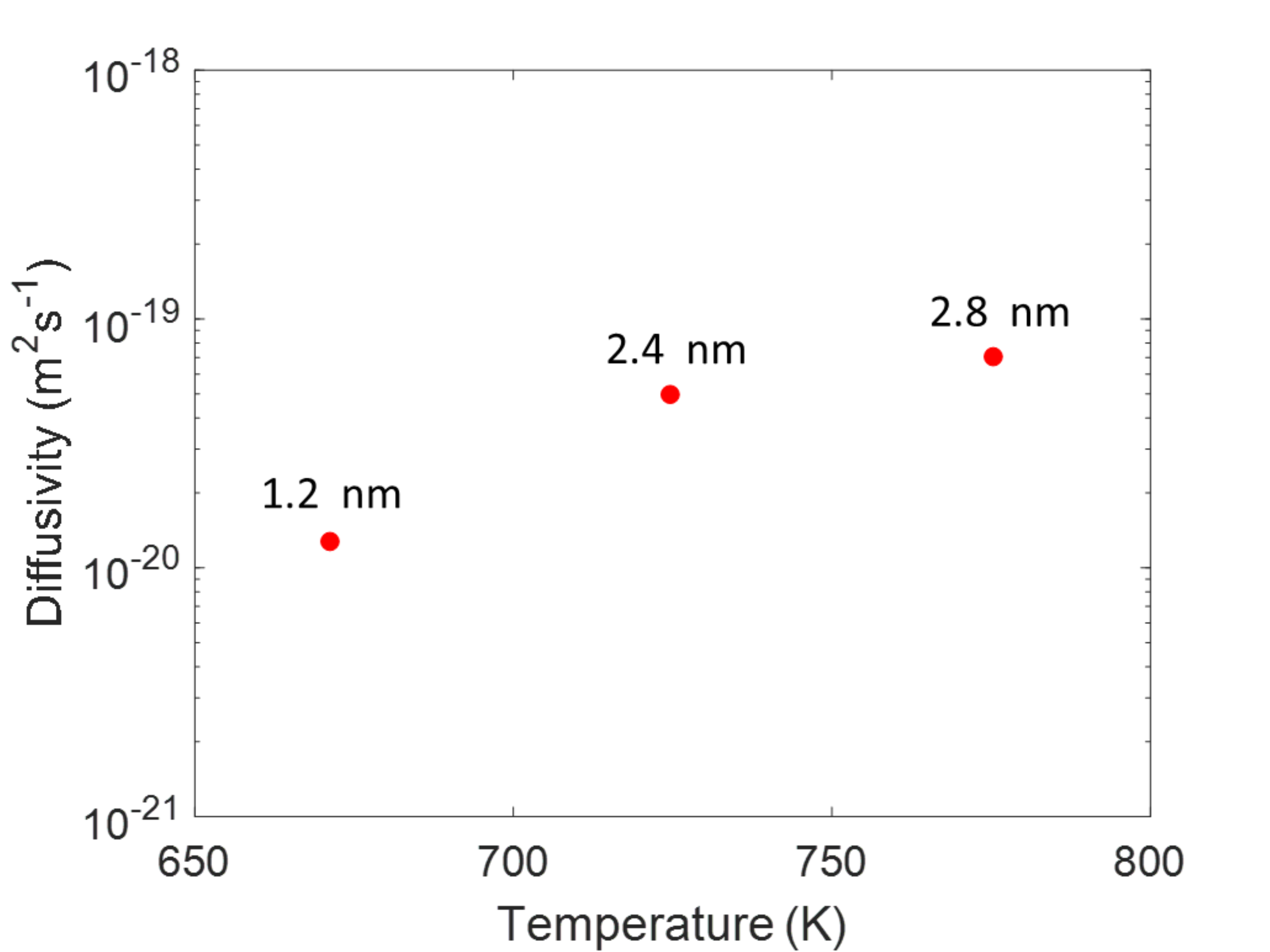}
	\caption{Atomic diffusivity in a-Si after 60 seconds of annealing at 673 K, 723 K, and 773 K \cite{straus_short_2016}. Numbers above each data point represent diffusion lengths at that temperature. }
	\label{fig:exp_diffusivity}
\end{figure}

Similar results were also experimentally observed recently in neutron reflectometry in \textsuperscript{29}Si/\textsuperscript{nat}Si isotope multilayers \cite{straus_self-diffusion_2016, straus_short_2016}. Isotope multilayers are stacks of 10 layers of [\textsuperscript{29}Si (5 nm) /\textsuperscript{nat}Si (16 nm)] deposited by ion-beam sputtering. In the neutron reflectometry, Bragg peaks are formed by the reflection of the neutrons at the isotope interfaces. Changes in the Bragg peak intensities at various temperatures compared to that of the as-deposited sample intensity leads to the self-diffusion measurements. This apparatus is sensitive to low diffusivities down to $10^{-25}$ m\textsuperscript{2} s\textsuperscript{-1} and very small diffusion lengths of 1 nm and below. Several temperature dependent diffusivity measurements under argon at ambient pressure during various time periods were done. The diffusivity measurements after 60 seconds of annealing at 673 K, 723 K, and 773 K are depicted in Fig. \ref{fig:exp_diffusivity} \cite{straus_short_2016}. Diffusion lengths, $d = (2Dt)^{1/2}$, are also represented on top of each data point. As the diffusivity is an ensemble averaged measurement, the distribution of the diffusion lengths for each atom is unknown, but it is apparent that on average atomic diffusion is quite significant in a-Si at these temperatures.

Quantitative comparison between experiments \cite{straus_short_2016} and our computations is difficult due to the large differences in simulation times (on the order of ps to ns) compared to experiments (minutes to hours). Nonetheless, it is evident from both experiments and calculations that the fundamental assumption in normal mode analysis that atoms vibrate around their equilibrium positions may be unsatisfied, especially at high temperatures but still below the amorphous silicon glass transition temperature of $\sim$ 1000 K \cite{hedler_amorphous_2004}. In contrast, normal mode analysis methods are routinely used for various amorphous solids at high temperatures up to 1200 K \cite{lv_non-negligible_2016,lv_examining_2016}.

\clearpage

\section{discussion}

Rather than considering propagating vibrations as individual normal modes (propagons), collective excitations characterized by the dynamic structure factor \cite{van_hove_correlations_1954} and wavepacket analysis \cite{beltukov_boson_2016, liu_triggering_2014, damart_nanocrystalline_2015, moon_propagating_2018, daly_picosecond_2004, daly_picosecond_2009, kim_origin_2021} may be perhaps more suitable to describe the propagating vibrations in amorphous solids where no assumption of normal modes is necessary. Inelastic scattering measurements and dynamic structure factor calculations from molecular dynamics are widely used to describe the collective excitations in numerous materials, ranging from liquids \cite{sette_collective_1995, sinn_coherent_1997, sinn_microscopic_2003} and solids \cite{sette_dynamics_1998, benassi_evidence_1996, masciovecchio_evidence_2006, baldi_damping_2017, monaco_anomalous_2009} to quantum gas \cite{landig_measuring_2015} and can provide crucial information such as group velocity and attenuation rates, vital to characterize thermal transport phenomena. Utilizing the collective excitations, recent works including our prior works \cite{kim_origin_2021} and others \cite{tlili_enhancement_2019} have been successful to describe thermal transport in glasses at a wide range of temperatures. In our view, normal mode analysis was instead historically utilized with extrapolation or model fitting schemes to characterize collective excitations from a simulation/theory standpoint due to the insufficiently large simulation domains to directly characterize the collective excitations by dynamic structure factor calculations as most amorphous solids have low Ioffe-Regel crossover frequencies ($\leq$ 1 THz) and wavevectors. It is also worth mentioning here that despite the extensive theoretical and numerical works on amorphous silicon, there had not been experimental measurements to directly characterize the collective excitations up to THz range in a wavevector and frequency resolved manner, possibly due to the large amount of samples required with a characteristic length of mm or cm required by typical inelastic scattering measurements. 

\begin{figure}
	\centering
	\includegraphics[width=0.9\linewidth]{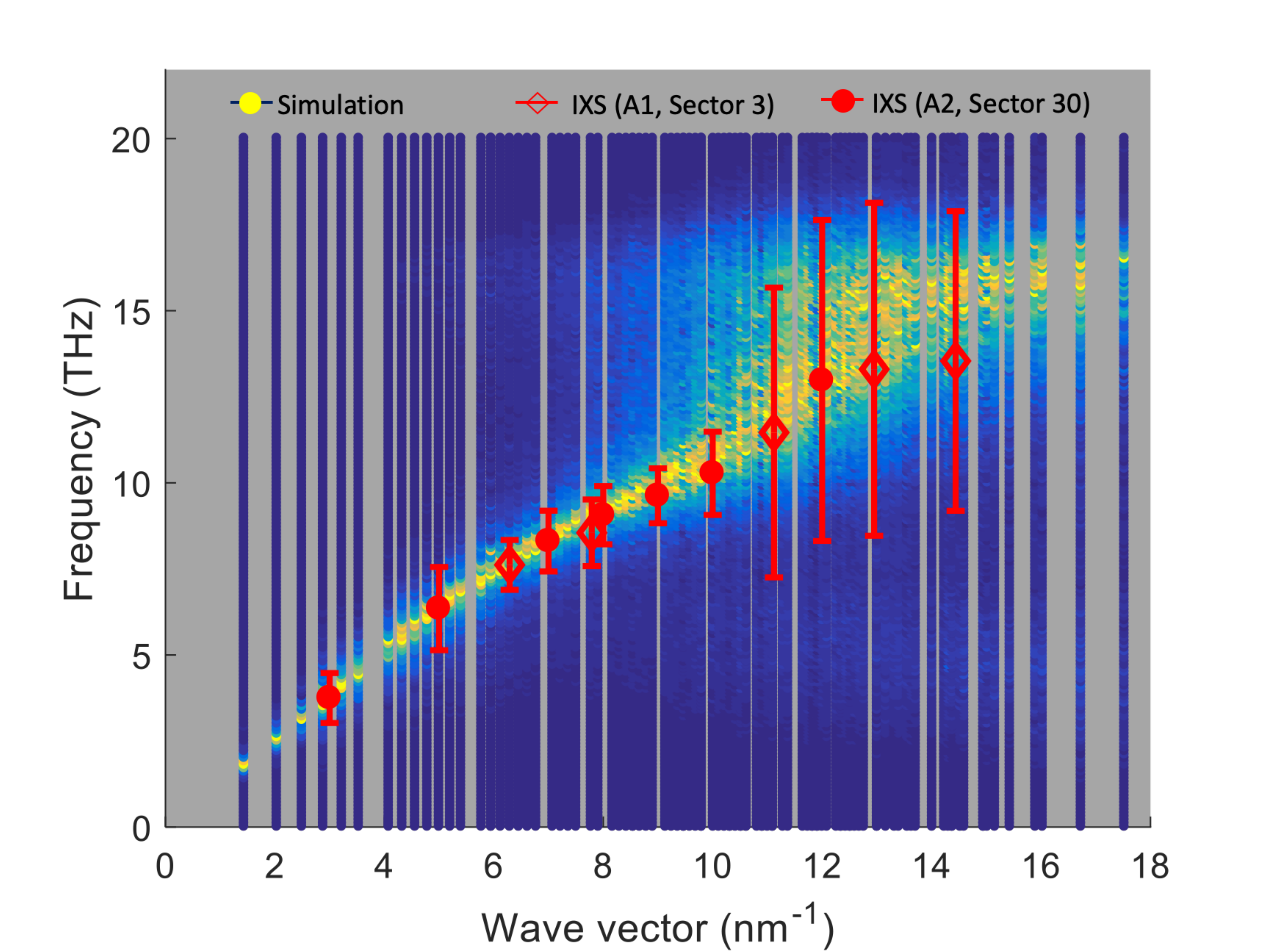}
	\caption{Dispersion relations from our molecular dynamics \cite{moon_propagating_2018} and inelastic x-ray scattering (IXS) \cite{moon_thermal_2019}. The IXS results are based on two different samples A1 and A2 at different beamlines HERIX sector 3 and 30, respectively at Advanced Photon Source. The errorbars in experimental data represent full widths at half maximum of inelastic peaks. Clear crisp phonon dispersion is observed up to 10 THz and good agreements between simulations and experiments are shown. Similar phonon dispersions have been also shown in other molecular dynamics works \cite{beltukov_boson_2016, larkin_thermal_2014}.}
	\label{fig:dispersion}
\end{figure}

Recently, we circumvented this sample quantity problem by powderizing a-Si films in argon gas environment to meet the length requirement and measured dynamic structure factor for the first time for a-Si \cite{moon_thermal_2019}. Our measurements \cite{moon_thermal_2019} and prior dynamic structure factor predictions \cite{beltukov_boson_2016, moon_propagating_2018, moon_thermal_2020} demonstrate consistent results that collective excitations or propagating acoustic excitations exist up to surprisingly high $\sim$ 10 THz with a clear dispersion rather than 2 to 3 THz predicted by other methods mentioned previously as shown in Fig. \ref{fig:dispersion}. In addition, there was a very weak temperature dependence in the inelastic peak widths in both experiments and molecular dynamics simulations, highlighting the weak anharmonicity, consistent with temperature dependent thermal conductivity measurements. It will, therefore, be interesting to see if the collective excitations from standard many-body physics and scattering theory can consistently and quantitatively describe thermal transport of propagating vibrations in other amorphous solids.

The theory of diffusons and locons has been widely used to describe thermal transport in various amorphous solids as propagating vibrations alone are not sufficient in describing the total thermal conductivity and in explaining the lack of size effects in thermal conductivity in many thin amorphous films and nanotubes \cite{wingert_thermal_2016}. Including both diffusons and propagons in analysis, thermal conductivity predictions can often match the experimental measurements well \cite{braun_size_2016, kwon_unusually_2017}. However, to the best of our knowledge, there has not been an explicit experimental proof of the Allen and Feldman theory of diffusons and locons such as mode specific thermal diffusivity measurements of diffusons. Hence, it remains an open question whether the theory of diffusons and locons are unique in describing non-propagating vibrations in amorphous solids. Perhaps, a new theory that does not depend on normal modes describing the non-propagating vibrations or new experimental techniques that explicitly characterize them will emerge in the future. 

As mentioned in the previous section, noticeable atomic diffusion can exist in amorphous solids, especially at high temperatures and it is becoming especially important for applications including glass electrolytes in solid-state batteries \cite{han_high_2019, grady_emerging_2020}. Rather than utilizing normal mode 
formalism, we propose a unified thermal conductivity model for liquids and glasses as $k_{total} = k_{collective} + k_{overdamped} + k_{diffusion}$ where $k_{collective}$ is from collective excitations, $k_{overdamped}$ is from overdamped vibrations, and $k_{diffusion}$ is due to atomic diffusion.  
The effect of atomic diffusion in thermal transport including its interactions with collective excitations and overdamped vibrations remains largely unexplored and these will be the subject of our next work.

\section{Conclusion}
Amorphous silicon has been a model glass to study thermal transport in glasses due to its relatively simple monatomic composition and industrial applications. Decades of research efforts utilizing normal mode methods in a-Si have been examined and the general consensus from these works has been discussed. A careful analysis of various works, however, clearly shows that there exist a number of discrepancies in the consensus, from results invalidating assumptions made in these methods to explicit inconsistencies when comparing to the measurements. Pinpointing the exact origin of these discrepancies is challenging but the common aspects that these methods share are the assumptions that (i) atoms vibrate around their equilibrium positions and that (ii) normal modes are the fundamental heat carriers in amorphous solids. We show through calculations that some atoms are prone to diffusion even at room temperature, contrasting the first assumption of normal modes mentioned above. %It may be the case that at sufficiently low temperatures or for some special atomic configurations, atoms do not have enough kinetic energy to overcome their local potential well and they do vibrate around their equilibrium positions. However, \textit{are normal modes good representations of atomic dynamics in the amorphous solid in general?} 
It is important to note that the degree of error that the atomic diffusion brings to the normal mode analysis is unexplored but an alternative scheme of studying atomic dynamics that does not rely on normal modes or assumptions of equilibrium positions is needed. Historically, atomic dynamics in amorphous solids was investigated using the ideas and concepts that were applied to crystals and recently, a unified theory of thermal transport of crystals and glasses has emerged \cite{simoncelli_unified_2019}. However, to generally consider thermal transport in amorphous at all temperatures, it is our view that we should consider amorphous solids as highly viscous liquids when studying their atomic dynamics and thermal transport. 

\section{Acknowledgement}

This work was supported by the US Department of Energy, Office of Science, Basic Energy Sciences, Materials Sciences and Engineering Division. We are grateful to Prof. Austin J. Minnich for helpful discussions over the years and we thank Dr. Lucas Lindsay for his comments on the manuscript and insightful discussions. The numerical calculations have been performed at Sherlock High Performance Computer Cluster at Stanford University. 

\section{Data availability}
The data that support the findings of this study are available from the corresponding author upon reasonable request.

\bibliography{My.bib}

\end{document}